\documentstyle[aps,prl,multicol]{revtex}

\begin{document}


\newcommand{\be}{\begin{equation}}
\newcommand{\ee}{\end{equation}}
\newcommand{\ba}{\begin{eqnarray}}
\newcommand{\ea}{\end{eqnarray}}


\title{Strong Enhancement of $\varepsilon'/\varepsilon$ Through Final State
  Interactions}


\author{ Elisabetta Pallante}
\address{
Facultat de F\'{\i}sica, Universitat de Barcelona, Diagonal 647,\\
E-08028 Barcelona, Spain, e--mail: pallante@ecm.ub.es }
\author{ Antonio Pich}
\address{ Departament de F\'{\i}sica Te\`orica, IFIC, 
Universitat de Val\`encia -- CSIC, \\ 
Apt. Correus 2085, E-46071 Val\`encia, Spain, e-mail: Antonio.Pich@uv.es }

\date{\today}
\maketitle

\begin{abstract}

We quantify the important effect of strong final state interactions in the 
weak $K\to2\pi$ amplitudes, using the measured $\pi$--$\pi$ phase shifts 
with $J=0$  and $I=0,2$. The final rescattering of the two pions provides a 
strong enhancement of the $\Delta I =1/2$ amplitude, which so far has been 
neglected in the theoretical predictions of $\varepsilon'/\varepsilon$.
This correction increases the Standard Model prediction of
$\varepsilon'/\varepsilon$ to values in good agreement with the 
experimental measurements. 

\end{abstract}



\begin{multicols}{2}

\narrowtext

It is well known that, at centre--of--mass
energies around the kaon mass, the strong S--wave $\pi$--$\pi$
scattering generates a large phase shift difference
$\left(\delta^0_0 - \delta^2_0\right)(m_K^2)= 45^\circ\pm 6^\circ$
between the $I=0$ and $I=2$ partial waves\cite{GM}. 
In the usual description of $K\to 2\pi$ decays,
this effect is explicitly taken into account,
through the following decomposition of the relevant isospin
amplitudes with $I=0,2$: 
\be\label{eq:AIdef}
{\cal A}_I \equiv A\left[ K\to (\pi\pi )_I\right] \equiv A_I \;
e^{i\delta^I_0} \, .
\ee
It has also been suggested \cite{TR88,IS90,KA91}
that final state interactions (FSI) play an
important role in the observed enhancement of the $I=0$ decay amplitude,
$A_0/A_2\approx 22.2$.   
However, their impact on the direct CP-violating parameter
$\varepsilon'/\varepsilon$ has never been properly estimated.

At lowest order in Chiral Perturbation Theory (ChPT), $O(p^2)$,
the decay amplitudes do not contain any strong phase:
\be\label{eq:chpt}
{\cal A}_I\big|_{O(p^2)} = 
- {G_F\over\sqrt{2}}\, V_{ud}^{\phantom{*}} V_{us}^* \: C_I \: 
\sqrt{2} f_\pi \: (m_K^2- m_\pi^2)  \, ,
\ee
where
$C_0 = g_8 + \frac{1}{9} g_{27}$ and
$C_2 = \frac{5}{9} \sqrt{2}\, g_{27}$,
with $g_8$ and $g_{27}$ unknown chiral couplings, corresponding to
the two lowest-order $\Delta S = 1$ operators in the momentum expansion,
transforming as $(8_L,1_R)$ and
$(27_L,1_R)$, respectively, under the chiral group
\cite{GPR86,PR91}.
The phenomenological determination of these couplings from
$K\to 2\pi$,
taking the measured phase shifts into account through (\ref{eq:AIdef}),
gives
$|C_0| \approx 5.1$ and $|g_{27}|\approx 0.29$.

The above procedure is not quite consistent, because the strong phases 
$\delta^I_0$ are put by hand. Those phases originate in the
final rescattering of the two pions and, therefore, are generated by
chiral loops which are of higher order in the momentum expansion.
Since the strong phases are quite large, one should expect large 
higher--order unitarity corrections.
The existing one--loop analyses of $K\to 2 \pi$ \cite{KA91,BPP,JHEP} 
show in fact that the pion
loop diagrams provide an important enhancement of the $A_0$ amplitude,
implying a sizeable reduction ($\sim 30\% $) of the fitted $|g_8|$
value. However, the phase-shift $\delta^0_0$ predicted  by the one--loop
calculation is still lower than its measured value, which indicates that
a further enhancement should be expected at higher orders.

Many attempts have been made to compute the amplitudes $A_I$ from
first principles 
\cite{GPR86,PR91,JP94,BBG87,Buras,KPR98,Lattice,Trieste,Dortmund,BP95}. 
Although those calculations have
provided encouraging results, we are still far from getting accurate
predictions. In many approaches FSI are not included
in the computational framework. This is the case of present lattice
calculations \cite{Lattice}, which are only able to compute the one pion
$\langle\pi|{\cal H}_{\Delta S=1}|K\rangle$ matrix elements, or
estimates at leading order in the $1/N_C$ expansion 
\cite{PR91,BBG87,Buras,KPR98}
(the phases $\delta^I_J$ are zero at leading order).
Other approaches \cite{BBG87,Trieste,Dortmund}
include some FSI effects, but in a
rather incomplete (and untrusty) way; however, the fact
that the values found for the $A_0$ amplitude are larger
than in the previous two methods is a clear
indication of the important role of FSI.

Unitarity and analyticity constraints permit a resummation of the effects due 
to FSI \cite{PLBFF}.
Let us consider the off-shell amplitudes ${\cal A}_I(s)$, with
$s = \left(k_{\pi_1} + k_{\pi_2}\right)^2$ the invariant mass squared
of the final two--pion state.
In the elastic region 
Watson final-state theorem relates 
the imaginary part of the amplitudes ${\cal A}_I(s)$ to the strong phases 
$\delta^I_{l=0}(s)$ of the partial wave amplitudes $T^I_{l=0}(s)$ which 
describe the rescattering of the two pions in the final state.
We can write a
1-subtracted dispersion relation for the amplitude ${\cal A}_I(s)$,
which has the following Omn\`es solution \cite{TR88,OM58}:
\ba\label{DISP}
{\cal A}_I(s) &=& A_I^{(0)} \; \left( s - m_\pi^2\right) \;\Omega_I(s)\, ,
\\ 
\Omega_I(s) & \equiv & \exp{\left\{ 
{(s-m_\pi^2)\over \pi}\int_{4m_\pi^2}^\infty\, 
{dz\over (z-m_\pi^2)} \, {\delta^I_0(z)\over (z-s-i\epsilon)}\right\} } .
\nonumber
\ea
We have imposed the subtraction at $s=m_\pi^2$, where the 
amplitude has a zero.
The constant $A_I^{(0)}$ is the first derivative of ${\cal A}_I(s)$ at
the subtraction point, i.e. the coefficient of the linear term in
$\left( s - m_\pi^2\right)$ of the Taylor expansion around $s= m_\pi^2$;
up to small higher--order mass corrections, 
it is given by the leading $O(p^2)$ term in the chiral expansion. 

The dispersion integral in eq.~(\ref{DISP}) has an imaginary part and a real 
part, so that the complete amplitude can be otherwise written as 
\begin{equation}
{\cal A}_I(s) = A_I^{(0)} \, \left( s - m_\pi^2\right) \,
e^{i\delta^I_0(s)}\,\Re_I(s) \, ,
\label{REAL}
\end{equation}
where $\Re_I(s)$ originates from the real part of the dispersion integral,
which is directly related to the strong phase shift. 
The factor $\Re_I(s)$ has never been taken into account in the 
short--distance calculation of the  amplitudes ${\cal A}_I(s)$. 

In this letter we show how the lacking of such term $\Re_I$  does affect 
numerically both the ratio $A_0/A_2$ and the direct CP violation parameter
$\varepsilon^\prime /\varepsilon$. The result that we obtain is the needed 
one to move both quantities towards the experimental findings.

The dispersion relation in eq.~(\ref{DISP}) is strictly valid only in the 
elastic region. Above the first inelastic threshold a coupled--channel 
analysis will be demanded \cite{PROGRESS}.
Here we limit ourselves to the very simple numerical analysis in the elastic 
region, which at $s=m_K^2$ should give a quite good approximation to the 
exact result, including inelastic effects.

We have used the simple parametrizations of the strong phases 
$\delta^I_0(s)$ given in ref.~\cite{SCHENK}.
They reproduce well the experimental data 
up to the $K\bar{K}$ threshold. 
Other recent parametrizations of the experimental data
\cite{MANY} performed coupled--channels analyses,
which we do not need in this simple first step calculation.  
The $I=2$ amplitude is found to be
elastic and well reproduced even up to 1.6 GeV.
We then evaluated numerically the dispersion integrals up to 1 
GeV for $I=0$ and 1.6 GeV for $I=2$.

Taking $s=m_{K}^2$, we find that the on-shell $K\to 2\pi$ decay
amplitudes get the dispersive correction factors:
\begin{equation}
\Re_0 =   1.41 \pm 0.06 \qquad , \qquad    
\Re_2 = 0.92\pm 0.02 \, .
\label{NUM_0}
\end{equation}
The central values have been obtained by using the 
parametrizations of $\delta^I_0(s)$ quoted as 
best fits in ref.~\cite{SCHENK}. 
To estimate the errors we have redone the numerical analysis with
two other sets of parametrizations, given in ref.~\cite{SCHENK},
which underestimate and overestimate, respectively, the phase shift data.

Since we have not included inelastic channels,
we have not taken into account contributions from energies above 1 GeV 
(1.6 GeV for $I=2$). This is a quite conservative attitude, since
those contributions tend to increase (decrease) $\Re_0$  ($\Re_2$) to
slightly larger (smaller) values.
The details of the numerical analysis will be given elsewhere \cite{PROGRESS}.

The corrections induced by FSI in the moduli of the decay amplitudes
${\cal A}_I$ have resulted in an additional enhancement of the
$\Delta I=1/2$ to $\Delta I=3/2$ ratio, 
\be\Re_0/\Re_2 = 1.53\pm 0.07 \, .\label{eq:ratio}\ee
This factor multiplies the enhancement already found at short distances.
This is a quite large correction, not taken into account previously,
which improves all existing calculations of $A_I^{(0)}$.
Taking the $\Re_I$ correction into account, the experimental
$A_I$ amplitudes imply the following corrected values for the
lowest--order $\Delta S=1$ chiral couplings in (\ref{eq:chpt}):
\be
|C_0| \approx 3.6 \qquad , \qquad
|g_{27}| \approx 0.32 \, .
\ee
This ``experimental'' numbers are not very far from the 
short--distance estimates obtained in the first of refs.~\cite{PR91}.

It is interesting to compare our results (\ref{NUM_0})
with the ones obtained by using the lowest--order estimate of the strong phases
$\delta^I_0(s)$ in ChPT.
We performed one step towards the unitarization of the amplitudes ${\cal A}_I(s)$ 
by using ($f_\pi\sim 93$ MeV) 
\begin{equation}
\tan{\delta^{0;2}_0(s)} = \sqrt{ 1-{4m_\pi^2\over s}}\;\left ( {2s-m_\pi^2\over
32\pi f_\pi^2}\,\, ;\,\, {2m_\pi^2-s\over 32\pi f_\pi^2}\right )\, ,
\label{CHPT}
\end{equation}
which gives the usual representation of $\delta^{0,2}_0(s)$ at threshold 
($s\sim 4m_\pi^2$), where $\tan{\delta^{0,2}_0}\sim \delta^{0,2}_0$.
For $I=0$ the lowest--order ChPT parametrization of eq. (\ref{CHPT})
integrated up to 1 GeV gives
$\Re_0 = 1.21$, sensitively underestimating the dispersion 
integral over the fitted phase as expected. 
For $I=2$ the situation is similar; lowest--order ChPT provides a good 
description of the strong phase only at low energies.
We obtain in this case $\Re_2 = 0.83$, also
lower than the value in eq. (\ref{NUM_0}) obtained using the fitted phase.
Note, however, that the ratio $\Re_0/\Re_2 = 1.46$, although slightly lower 
than the value (\ref{eq:ratio}), stays within the quoted error bar.
This is in agreement with the observation made in ref.~\cite{GM} that
the difference $\delta^0_0 - \delta^2_0$ gets smaller chiral corrections
than the individual phase shifts.

\vspace{0.5cm}
{\bf Implications for $\varepsilon'/\varepsilon$.---}

The most striking consequence of the correction factors $\Re_{0,2}$
is a sizeable modification of the numerical short--distance
estimates for the direct CP--violation 
parameter $\varepsilon^\prime/\varepsilon$. 
A handful way of writing this quantity,
used in all theoretical short--distance calculations up to date, can be as 
follows \cite{Buras}
\begin{equation}
{\varepsilon^\prime\over\varepsilon} = 
\mbox{Im}\,\lambda_t\cdot e^{i\Phi}\cdot
\left [P^{(1/2)} - P^{(3/2)}\right ]\, ,
\label{EPS}
\end{equation}
where 
$\lambda_t =  V_{ts}^* V_{td}^{\phantom{*}}$,
the phase $\Phi = \Phi_{\varepsilon^\prime} -  \Phi_\varepsilon \simeq 0$ 
and the 
quantities $P^{(1/2)}$ and $ P^{(3/2)}$ contain the contributions from the 
hadronic matrix elements of four-quark operators 
with $\Delta I =1/2$ and $3/2$ respectively:
\begin{eqnarray}
P^{(1/2)}&=& r \,\sum_i y_i(\mu)\, \langle Q_i(\mu)\rangle_0 \,
(1-\Omega_{\eta+\eta^\prime}) \, ,
\nonumber\\
 P^{(3/2)}&=& {r\over \omega} \,\sum_i y_i(\mu)\, \langle Q_i(\mu)\rangle_2 
\, . \label{P_I}
\ea
Here, $\langle Q_i\rangle_I\equiv \langle (\pi\pi )_I\vert Q_i\vert K\rangle$,
 $r$ and $\omega$ are given by
\be
r= {G_F\over 2 \vert\varepsilon\vert}\, {\omega\over \mbox{Re}A_0}\, 
\qquad , \qquad
\omega = {\mbox{Re}A_2\over \mbox{Re}A_0}\, ,
\label{eq:r_omega}
\ee
and the parameter
\be
\Omega_{\eta+\eta^\prime} = {1\over \omega} 
{(\mbox{Im}A_2)_{\mbox{\small{I.B.}}}
\over \mbox{Im}A_0}
\label{eq:isospin}
\ee
parametrizes isospin breaking corrections. It is usually set to 
$\Omega_{\eta+\eta^\prime}\approx 0.25$ \cite{Omega} with large uncertainties.

The Wilson coefficient factors $y_i(\mu)$ have been computed to next-to-leading
order accuracy \cite{Buras,Lattice}.
Since the hadronic matrix elements are quite uncertain theoretically,
the CP--conserving amplitudes $\mbox{Re}A_I$, and thus
the factors $r$ and $\omega$, are set to their experimentally
determined values; this automatically includes the FSI effect.
All the rest in the numerator is {\em theoretically} predicted via
short--distance calculations, because the leading contributions come from the
operators $Q_6$ and $Q_8$ whose matrix elements cannot be directly
measured from the $K\to 2\pi$ decay rates.

As a consequence, since the relevant matrix elements $\langle Q_{6,8}\rangle_I$
are usually taken from lattice calculations \cite{Lattice} or large--$N_C$
estimates \cite{Buras}, which do not include FSI corrections,
this procedure produces a mismatch with the FSI 
included phenomenologically in the values of $r$ and $\omega$.
This can be easily corrected, introducing in the numerator
the dispersion factors $\Re_I$ that we have estimated.
This implies a large enhancement of the predicted value of
$\varepsilon'/\varepsilon$ by roughly a factor of 2.

A fast way to estimate the numerical enhancement is through the
approximate formula \cite{Buras}
\be
{\varepsilon'\over\varepsilon} \sim
\left [ B_6^{(1/2)}(1-\Omega_{\eta+\eta^\prime}) - 0.4 \, B_8^{(3/2)}
 \right ]\, ,
\label{EPSNUM}
\ee
where $B_6^{(1/2)}$ and $B_8^{(3/2)}$ parametrize the matrix elements
of the QCD penguin operator $Q_6$ and the electroweak penguin operator $Q_8$, 
respectively, in units of their vacuum insertion approximation.
These parameters are usually taken to be
(from Lattice calculations \cite{Lattice} and $1/N_C$ considerations 
\cite{Buras})  
$B_6^{(1/2)}= 1.0\pm 0.3$ and $B_8^{(3/2)} = 0.8\pm 0.2$.
Since those estimates do not include the FSI effect, their values should be
multiplied by the appropriate factors $\Re_0$ and $\Re_2$, 
respectively\cite{footnote}.
However, the term $B_6^{(1/2)}\Omega_{\eta+\eta^\prime}$ in eq. (\ref{EPSNUM})
should be multiplied by  $\Re_2$ and not by $\Re_0$,
because it corresponds to two final pions with $I=2$.

The three terms of eq. (\ref{EPSNUM})
should then be corrected to 
$B_6^{(1/2)}|_{FSI}= 1.4\pm 0.3$, $B_8^{(3/2)}|_{FSI} = 0.7\pm 0.2$ and 
$B_6^{(1/2)}\Omega_{\eta+\eta^\prime}|_{FSI}= 0.23\pm 0.07$. In the latter,
we have set $\Omega_{\eta+\eta^\prime}$ equal to 0.25,
disregarding its (probably large) error.
This amounts to a FSI enhancement of $\varepsilon'/\varepsilon$ 
by a factor 2.1.

Thus, the so-called ``central'' value in refs.~\cite{Buras},
$\varepsilon'/\varepsilon = 7.0\times 10^{-4}$ gets increased by
the FSI correction to
$\varepsilon'/\varepsilon = 15\times 10^{-4}$,
which compares well with the present experimental world average
\cite{exp}
\be\label{eq:exp}
\mbox{Re}\left(\varepsilon'/\varepsilon\right) = 
(21.2\pm 4.6)\times 10^{-4} \, .
\ee

The final theoretical prediction of $\varepsilon'/\varepsilon$
depends on other hadronic and
quark mixing parameters, which introduce a rather large uncertainty.
It has been said in refs.~\cite{Buras} and \cite{Lattice} that
the theoretical prediction can only be made consistent with
the experimental value (\ref{eq:exp}) for rather extreme values
of the input parameters, requiring a conspiracy of several inputs
in the same direction.
The important point we want to stress here is that, once the effect
of FSI has been properly taken into account, the experimental value
of $\varepsilon'/\varepsilon$ can be easily obtained with reasonable
values \cite{PP95} of the different inputs.

Our calculation of the unitarity $\Re_I$ correction factors,
provides a hint on why the Trieste \cite{Trieste}
and Dortmund \cite{Dortmund} groups were able to obtain
larger values of $\varepsilon'/\varepsilon$ than the Munich \cite{Buras}
and Rome \cite{Lattice} collaborations. 
The first two groups did include some FSI
corrections, at the one-loop level, 
which pushed their predictions in the right direction.
The use of the dispersive factors $\Re_I$ offers a more powerful 
and consistent way
to estimate these FSI effects, because the Omn\`es exponential
$\Omega_I(s)$ sums the unitarity logarithms to all orders in the
chiral expansion.

 More work is still needed in order to get an accurate prediction for 
$\varepsilon'/\varepsilon$. In the meanwhile, our
calculation demonstrates that, within the present uncertainties,
the Standard Model
theoretical estimates can easily accommodate the experimental
value, without any need to invocate a new
physics source of CP violation phenomena.

\vspace{0.5cm}
This work has been supported in part by the European Union TMR Network
EURODAPHNE (Contract No. ERBFMX-CT98-0169), and by DGESIC (Spain) under
grant No. PB97-1261. EP is supported by the Ministerio de Educaci\'on 
y Cultura (Spain).

\hspace{-2cm}

\end{multicols}

\begin{thebibliography}{99}
\bibitem{footnote}
The electroweak penguin operator $Q_8$ contributes to $K\to\pi\pi$
at $O(p^0)$ in ChPT. Thus, the CP--odd amplitude
$\mbox{\rm Im} A_2$ does not
vanish at $s=m_\pi^2$ and the corresponding Omn\`es solution is modified
in a trivial way. One gets the same Omn\`es exponential 
$\Omega_I(s)$ as in eq.~(\ref{DISP}) and
the same dispersive factor $\Re_2$ than for the
CP--even amplitude $\mbox{\rm Re} A_2$.
\end{thebibliography}
\end{document}